\documentclass[aps,pra,twocolumn,amsmath,amssymb,groupaddresses,superscriptaddress,showpacs]{revtex4}
\usepackage{times}
\usepackage{latexsym}
\usepackage{graphicx}
\usepackage{amsmath,amssymb}
\usepackage{verbatim,bbm}
\newtheorem{theorem}{Theorem}

\begin{document}

\title{Quantum-locked key distribution at nearly the classical capacity rate}

\author{Cosmo Lupo}
\affiliation{Research Laboratory of Electronics, Massachusetts Institute of Technology, Cambridge, MA 02139, USA}

\author{Seth Lloyd}
\affiliation{Research Laboratory of Electronics, Massachusetts Institute of Technology, Cambridge, MA 02139, USA}
\affiliation{Department of Mechanical Engineering, Massachusetts Institute of Technology, Cambridge, MA 02139, USA}

\begin{abstract}
Quantum data locking is a protocol that allows for a small secret key
to (un)lock an exponentially larger amount of information, hence yielding the strongest 
violation of the classical one-time pad encryption in the quantum setting.
This violation mirrors a large gap existing between two security criteria for quantum cryptography
quantified by two entropic quantities: the Holevo information and the accessible information.
We show that the latter becomes a sensible security criterion if an upper bound on the
coherence time of the eavesdropper's quantum memory is known.
Under this condition we introduce a protocol for secret key generation through a memoryless qudit channel.
For channels with enough symmetry, such as the $d$-dimensional erasure and depolarizing channels,
this protocol allows secret key generation at an asymptotic rate as high as the classical capacity minus one bit.
\end{abstract}


\maketitle

{\it Introduction.--} A famous theorem of Shannon's assesses the security of one-time pad encryption, and shows that
the secure encryption of a message of $n$ classical bits requires a key of at least $n$ bits~\cite{Shannon}.
When the message is encrypted in quantum bits or qubits, by contrast, the phenomenon of quantum
data locking (QDL)~\cite{QDL,CMP,Buhrman,Leung,Fawzi,Dupuis} shows that the key required for secure encryption
of an $n$ bit message can be much less than $n$.
In a typical QDL protocol, the legitimate parties, Alice and Bob, publicly
agree on a set of $N = MK$ codewords in a high-dimensional quantum system.  
From this set, they then use a short shared private key of $\log K$ bits to select a set
of $M$ codewords that they will use for sending information.
In the strongest QDL protocols known up to now, a key of {\it constant} length of about
$O\left(\log{1/\epsilon}\right)$ bits allows one to encrypt a message of $n$ bits, in such a way 
that if an eavesdropper Eve intercepts and measures the quantum system, then she cannot access more 
than about $\epsilon n$ bits of information about the message~\cite{Fawzi,phase}.

%%%

A number of works have been devoted to the role of QDL in physics and information theory~\cite{bh,CMP,Buhrman,Leung,Fawzi,Dupuis,phase,QEM,PRX}.
However, only recently has QDL been considered in the presence of noise.
Following the idea of the ``quantum enigma machine''~\cite{QEM} for applying QDL to cryptography,
a formal definition of the {\it locking capacity} of a communication channel has been recently
introduced in~\cite{PRX}, as the maximum rate at which information can be reliably and
securely transmitted through a (noisy) quantum channel. Unlike the private capacity 
(which requires the communication to be secure according to the Holevo information criterion), the
locking capacity requires security according to the {\it accessible information} criterion, possibly
with the assistance of a preshared secret key whose length grows sublinearly in the number
of channel uses.
Since the Holevo information is an upper bound on the accessible information, the locking capacity
is always larger than or equal to the private capacity. Clearly, the locking capacity cannot exceed
the classical capacity (that is, the maximum rate for classical communication without any privacy).
Two notions of capacity were defined in~\cite{PRX}: 
the {\it weak} locking capacity is defined by requiring security against an eavesdropper who measures 
the output of the complementary channel to the channel from Alice to Bob 
(that is, she measures the environment of the channel); 
the {\it strong} locking capacity is instead defined by assuming that
the eavesdropper is able to measure the very input of the channel.
In general, the weak locking capacity is larger than or at most equal to the strong locking capacity,
as any strong locking protocol also defines a weak locking one.
As shown in~\cite{AW}, there exist qudit channels with low ($1$ bit per channel use) or even zero private 
capacity whose weak locking capacity is larger than $\frac{1}{2}\log{d}$. 
In particular, the examples in~\cite{AW} refer to effectively noiseless channels whose 
classical capacity is $\log{d}$ bits.

Here we introduce a protocol that allows high rate QDL over a memoryless (noisy) qudit channel,
and we apply it to define a secret key generation protocol which is secure in the sense of strong
locking. 
The protocol allows secret key generation at a rate as high as the classical capacity minus one bit,
independently of the channel having any private capacity.
This result shows that by using a weaker security criterion (the accessible information) one can
increase the secret key generation rate up to almost the classical capacity.
As explained below, the accessible information becomes a sensible criterion in a scenario where
Alice and Bob know an upper bound on the coherence time of Eve's quantum memory.

{\it Overview.--} One of the most profound implications of QDL in quantum information theory is the
existence of a potentially large gap between two security criteria for quantum cryptography~\cite{Renner}.
Suppose that Eve has access to the state $\rho_{E|x}$ given that the 
classical message $x$ has been sent by Alice to Bob.
The widely accepted security criterion in quantum cryptography requires that Eve's state 
is $\epsilon$-close to being a product state in the operator trace norm~\cite{Renner}, that is,
\begin{equation}
\left\| \sum_x p_X(x) |x\rangle\langle x| \otimes \rho_{E|x} - \sigma \otimes \rho_E \right\|_1 
\leq \epsilon \, , \label{opnorm}
\end{equation}
where $\| \cdot \|_1 = \mathrm{Tr}|\cdot|$,
$p_X(x)$ is the probability that the input random variable $X$ takes value $x$, $\sigma = \sum_x p_X(x) |x\rangle\langle x|$,
and $\rho_E = \sum_x p_X(x) \rho_{E|x}$.
By application of the Alicki-Fannes inequality~\cite{Fannes}, Eq.~(\ref{opnorm}) implies 
\begin{equation}
\chi\left( \mathcal{E} \right) \leq 4 \epsilon \log{|X|} + 2 h_2(\epsilon)\, ,
\end{equation}
where $\chi\left( \mathcal{E} \right) := S\left(\rho_E\right) - \sum_x p_X(x) S\left(\rho_{E|x}\right)$ is Holevo information 
of the ensemble of quantum states $\mathcal{E} = \{ p_X(x) , \rho_{E|x} \}$, 
$S(\rho) := - \mathrm{tr}\rho\log{\rho}$ denotes the von Neumann entropy, 
$|X|$ is the cardinality of the input variable $X$,
and $h_2(\epsilon) = - \epsilon \log{\epsilon} - (1-\epsilon) \log{(1-\epsilon)}$
denotes the binary Shannon entropy.
A fundamental feature of the Holevo information is that it obeys the property of {\it total proportionality}~\cite{QDL}.
This means that if Eve is given $k$ bits (or $k/2$ qubits) of side information 
about the message, then her Holevo information cannot increase by more than $k$ bits.

%%%

In the early days of quantum cryptography, the accessible information criterion 
was used instead of the Holevo information (see, e.g.,~\cite{QKD}).
This criterion requires that the result of any measurement Eve can make on her share of the
quantum state is $\epsilon$-close to being uncorrelated with the message.
Suppose that a measurement $\mathcal{M}_{E \to \hat X}$ maps $\rho_{E|x}$ into the
classical variables $\hat X$ with conditional probability distribution $p_{\hat X | X}$. 
Then one considers the norm
\begin{align}
& \sup_{\mathcal{M}_{E \to \hat X}} \left\| p_{\hat X | X} p_X - p_{\hat X} p_X \right\|_1 := \nonumber \\
& \sup_{\mathcal{M}_{E \to \hat X}} \sum_{x, \hat x} \left| p_{\hat X | X}(\hat x | x ) p_X(x) - p_{\hat X}(\hat x) p_X(x) \right| \, , \label{vecnorm}
\end{align}
where $p_{\hat X}(\hat x) = \sum_x p_{\hat X | X}(\hat x | x ) p_X(x)$.
If~(\ref{vecnorm}) is less than $\epsilon$, then the Alicki-Fannes inequality implies~\cite{NOTA02}
\begin{equation}
I_\mathrm{acc}\left( \mathcal{E} \right) \leq 4 \epsilon \log{|X|} + 2h_2(\epsilon) \, ,
\end{equation}
where $I_\mathrm{acc}\left( \mathcal{E} \right) := \sup_{\mathcal{M}_{E \to \hat X}} I(X;\hat X)$ is the 
accessible information of the ensemble $\mathcal{E} = \{ p_X(x) , \rho_{E|x} \}$,
$I(X;\hat X) = H(X) + H(\hat X) - H(X\hat X)$ is the classical mutual information
between the message variable $X$ and the measurement result $\hat X$, and 
$H(X) = - \sum_x p_X(x) \log{p_X(x)}$ denotes the Shannon entropy.
Unlike the Holevo information, the accessible information does not obey the property of
total proportionality~\cite{QDL}. This implies that the accessible information is, in general, not stable under loss 
of information to Eve. That is, if Eve obtains $k$ bits of side information 
about the message there is no guarantee that her accessible information will increase by a proportionate amount
(and indeed it can increase by an arbitrarily large amount according to the QDL effect).

%%%

While it is clear that at a certain point Eve has to measure her share of the quantum state,
the accessible information criterion is sensitive to the time at which such a measurement takes place.
If Eve obtains a small amount of side information {\it before} she measures her share, then she could use this 
information to increase her accessible information by a disproportionate amount.
As a consequence, accessible information security is not, in general, composable~\cite{Renner},
that is, a protocol that is secure according to the accessible information criterion may not
remain so when used as a subroutine of another communication protocol.
On the other hand, if Eve obtains $k$ bits of side information {\it after} the measurement, then 
(since the classical mutual information obeys total proportionality) her accessible information 
cannot increase by more than $k$ bits and composable security will be granted \cite{NOTA-Holevo}.

As is customary in quantum key distribution, our secret key generation protocol is divided in two parts.
The first part is a QDL protocol in which Alice encodes her share of the raw key into 
quantum states and sends them to Bob via an insecure quantum channel.
After Bob measures the output of the channel he obtains his own share of the raw key that 
has to be reconciled with Alice's one.
The security of this part of the protocol is granted by the QDL effect and is
quantified by the accessible information.
In the second part of the protocol Alice sends error correcting information to Bob 
through a public channel (in our case there is no need for privacy amplification since 
the raw key is already secure due to QDL \cite{NOTA-nopa}). 
We are hence in a situation where the QDL protocol is used as subroutine of the
key distribution protocol. This implies that the latter will be secure only if the
former is secure in the composable sense. 
As discussed above, this is, in general, true only under the assumption that Eve has 
already measured her share of the quantum state when the second part of the protocol 
takes place.
If Alice knows that Eve's quantum memory has a coherence time not larger than $\tau$, 
then she can simply wait for a sufficiently long time before sending error 
correcting information to Bob through the public channel.
After such a time Eve has either made a measurement or her quantum memory has
completely decohered.
In both cases the security of the QDL protocol will be composable.

For any value of $\tau$ Alice and Bob can apply a doubly-blocked communication protocol, 
where they first send a data packet down the quantum channel, and then wait a time $\tau$ 
before doing all the required classical post-processing.
In the meantime Alice can keep sending Bob independent data packets that will be
processed at a later time.
The larger $\tau$ is, the longer Alice and Bob have to wait to guarantee the security of the protocol.
Clearly, too large values of $\tau$ would make the protocol unpractical. 
However, it is worth remarking that from an abstract point of view, in a stationary regime 
the asymptotic communication is independent of $\tau$ and it remains finite even 
in the limit $\tau \to \infty$.

%%%

{\it Accessible information security.--} Our starting point is a new QDL protocol 
defined for a memoryless $d$-dimensional channel (for any $d \geq 3$).
Upon $n$ uses of the qudit channel $\mathcal{N}$, the protocol allows one to lock 
classical information using an ensemble of input codewords $\mathcal{E}$ that are 
separable among different channel uses.
The protocol requires Alice and Bob to initially share a secret key of $\log{K_n}$
bits, which is consumed at an asymptotic rate of $\lim_{n\to\infty} \frac{1}{n} \log{K_n} = 1$ bit per channel use.

Let us fix a qudit basis $\{ |\omega\rangle \}_{\omega=1,\dots,d}$ 
and its Fourier conjugate $\{ |m\rangle \}_{m=1,\dots,d}$, with
\begin{equation}\label{FT}
|m\rangle = \frac{1}{\sqrt{d}} \, \sum_{\omega=1}^d e^{i 2\pi m \omega/d} |\omega\rangle \, .
\end{equation}
We consider the ``phase ensemble'' of qudit unitary transformations of the form:
\begin{equation}\label{Uphase}
U = \sum_{\omega=1}^d e^{i\theta(\omega)} |\omega\rangle \langle \omega| \, ,
\end{equation}
where the angles $\theta(\omega)$, for $\omega=1,\dots,d$, are $d$ i.i.d.\ random variables. 
We require that these variables are distributed in such a way
that $\mathbb{E}[e^{i\theta(\omega)}] = 0$~\cite{NOTA01}.
To define the QDL protocol upon $n$ uses of the channel, 
Alice and Bob publicly agree on a set of $K_n$ $n$-qudit unitaries of the form
$\{ \otimes_{j=1}^n U^j_k \}_{k=1,\dots,K_n}$. 
The value of the index $k$ plays the role of a secret key of $\log{K_n}$ bits 
initially shared by Alice and Bob.
Alice prepares with equal probability one of the
$d^n$ orthogonal vectors $|\boldsymbol{m}\rangle = \otimes_{j=1}^n |m^j\rangle$
(the $n \log{d}$ bits string $\boldsymbol{m}$ will serve as a raw key for Alice), and
then {\it scrambles} it by applying one of the unitary transformations, yielding
\begin{equation}
|\Psi_{\boldsymbol{m}k}\rangle = \otimes_{j=1}^n U^j_k |m^j\rangle 
=  \sum_{\boldsymbol{\omega}} \frac{ e^{i\sum_{j=1}^n \left[ 2\pi m^j \omega^j/d + \theta^j_k(\omega^j) \right]} }{\sqrt{d^n}} |\boldsymbol{\omega}\rangle \, .
\end{equation}

We first prove that if Eve (who does not know the value of the index $k$) 
intercepts the whole train of qudit systems and measures them,
then she can only retrieve a negligible amount of information about the input variable $\boldsymbol{m}$.
In particular, we show that there exist choices of the scrambling unitaries $U^j_k$ 
that guarantee that Eve's accessible information is arbitrarily small if $n$ is large enough.
To prove this, we show that this property is almost certainly true if
each $U^j_k$ is sampled i.i.d.\ from the phase ensemble of unitaries \cite{NOTA-proof}.

Let Eve intercept and measure the train of $n$ qudits sent by Alice. A measurement is described by a collection of POVM elements
$\{ \mu_i |\Phi_i\rangle\langle\Phi_i| \}_i$, where $\sum_i \mu_i = d^n$, $\mu_i >0$ and $|\Phi_i\rangle$ are unit
vectors (possibly entangled over the $n$ qudit systems).
Since Eve does not have access to the secret key, we have to compute the accessible information of the 
ensemble of states $\mathcal{E} = \{ p_{\boldsymbol{m}} , \frac{1}{K_n} \sum_{k=1}^{K_n} |\Psi_{\boldsymbol{m}k}\rangle \langle \Psi_{\boldsymbol{m}k} |\}$,
averaged over the values of the secret key,
where $p_{\boldsymbol{m}} = 1/d^n$ is the probability of the message $\boldsymbol{m}$.
A straightforward calculation then yields
\begin{equation}
I_{acc}\left( \mathcal{E} \right) = \log{d^n} - \min_{\{ \mu_i |\Phi_i\rangle\langle\Phi_i| \}} \sum_i \frac{\mu_i}{d^n} \, H[Q(\Phi_i)]  \, ,
\end{equation}
where $Q(\Phi)$ denotes the $d^n$-dimensional real vector with non-negative entries
\begin{equation}
Q_{\boldsymbol{m}}(\Phi) = \frac{1}{K_n} \sum_{k=1}^{K_n} |\langle \Phi | \Psi_{\boldsymbol{m}k} \rangle |^2 \, ,
\end{equation}
and 
\begin{equation}
H[Q(\Phi)] = - \sum_{\boldsymbol{m}} Q_{\boldsymbol{m}}(\Phi) \log{Q_{\boldsymbol{m}}(\Phi)}
\end{equation}
is its Shannon entropy
(notice that $\sum_{\boldsymbol{m}} Q_{\boldsymbol{m}}(\Phi) = 1$).

Since $\sum_i \mu_i/d^n = 1$, the positive coefficients $\mu_i/d^n$ can be interpreted as probability
weights. We can then apply a standard convexity argument (the minimum is never larger than the average) 
to obtain an upper bound on Eve's accessible information:
\begin{equation}\label{Iaccnl}
I_{acc}\left( \mathcal{E} \right) \leq \log{d^n} - \min_{|\Phi\rangle} \, H[Q(\Phi)]  \, ,
\end{equation}
where the minimum is over all $n$-qudit unit vectors.
According to this expression, an upper bound on the accessible information follows
from a lower bound on the minimum Shannon entropy $\min_{|\Phi\rangle} \, H[Q(\Phi)]$.

To show that $I_{acc}\left( \mathcal{E} \right)$ can be made arbitrarily small, we apply concentration
inequalities \cite{AM,Chernoff} to the quantities $Q_{\boldsymbol{m}}(\Phi)$'s. Notice that the latter
are random variables if the unitaries $U_k^j$ are chosen randomly from the 
phase ensemble.
The main idea is that the $Q_{\boldsymbol{m}}(\Phi)$'s will concentrate around their
mean value $1/d^n$. We prove (see \cite{suppmat})
that the probability of a deviation larger than $\epsilon/d^n$ is exponentially suppressed.
This property will be used to show that $I_{acc}\left( \mathcal{E} \right) \lesssim \epsilon \log{d^n}$ (up to
a probability exponentially small in $d^n$).
In order for this to be true, the number of different scrambling unitaries has to satisfy \cite{NOTA-key}
\begin{equation}\label{skey}
K_n > 2^{n+1} \left( \frac{1}{\epsilon^2} \ln{d^n} + \frac{2}{\epsilon^3} \log{\frac{5}{\epsilon}} \right) \, .
\end{equation}
This implies an asymptotic secret key consumption rate of 
$\lim_{n \to \infty} \frac{1}{n}\log{K_n} = 1$ bit per channel use.
We remark that we can put $\epsilon = 2^{-n^c}$, for any $c<1$, and still lock data with a
secret key consumption rate of $1$ bit independently of $d$.

%%%

{\it Secret key generation.--} As an example, we consider the case of a collective attack
by Eve, which induces the memoryless qudit channel $\mathcal{N}$ from Alice to Bob.
(Since our QDL is secure in the strong locking sense, it will be secure also in the
case of general coherent attacks.)
For any given value of $k$ Bob receives one of 
the $d^n$ equiprobable $n$-qudit states $\mathcal{N}^{\otimes n}\left(|\Psi_{\boldsymbol{m}k}\rangle\langle\Psi_{\boldsymbol{m}k}|\right)$
at the output of the channel.
For the sake of simplicity we consider the case of unitarily covariant channels, that is, 
satisfying $\mathcal{N}(U \rho U^\dag) = U \mathcal{N}(\rho)U^\dag$ for any qudit unitary $U$.
(For example, this is the case of the erasure and depolarizing channels.)
To decrypt the message Bob can apply the inverse unitary $\otimes_{j=1}^n {U^j_k}^{-1}$.
After the decryption, Bob obtains $n$ independent instances of the qudit ensemble of output
states $\{ 1/d , \mathcal{N}(|m\rangle\langle m|) \}$.
To decode, Bob applies a measurement on these states, obtaining a raw key $\hat{\boldsymbol{m}}$
given by the measurement outcomes.
Finally, to distill a perfectly correlated key Alice should send error correcting 
information to Bob. If Bob makes the optimal measurement, they will asymptotically achieve 
about $n \chi_\mathcal{N}(\mathcal{E})$ bits of common randomness, where
$\chi_\mathcal{N}(\mathcal{E}) = S[\frac{1}{d} \sum_m \mathcal{N}(|m\rangle\langle m|)] - \frac{1}{d}\sum_m S[\mathcal{N}(|m\rangle\langle m|)]$
is the Holevo information of the channel~\cite{HSW}.
At this stage we make use of the assumption that Alice knows an upper bound $\tau$ 
on the coherence time of Eve's quantum memory.
Since the error correcting information will be transmitted on a public communication channel, 
Alice must wait for a time larger than $\tau$ before being able to
safely send error correcting information to Bob.
In this way Alice and Bob establish a secret key of about $n \chi_\mathcal{N}(\mathcal{E})$
bits starting from one of about $n$ bits.
If $\chi_\mathcal{N}(\mathcal{E}) > 1$, they can then run the protocol again by recycling 
part of the obtained secret key and achieve an overall asymptotic rate of secret key generation of
$R = \chi_\mathcal{N}(\mathcal{E}) - 1$ bits per channel use.

In particular, for a unitarily covariant channel, such as
the qudit erasure channel and the qudit depolarizing channel,
the Holevo information $\chi_\mathcal{N}(\mathcal{E})$ equals the classical capacity $C_\mathcal{N}$:
hence, QDL allows for a secret key generation rate of $R = C_\mathcal{N} - 1$ bits,
just one bit below the channel classical capacity.

\begin{figure}[t]
\centering
\includegraphics[width=0.235\textwidth]{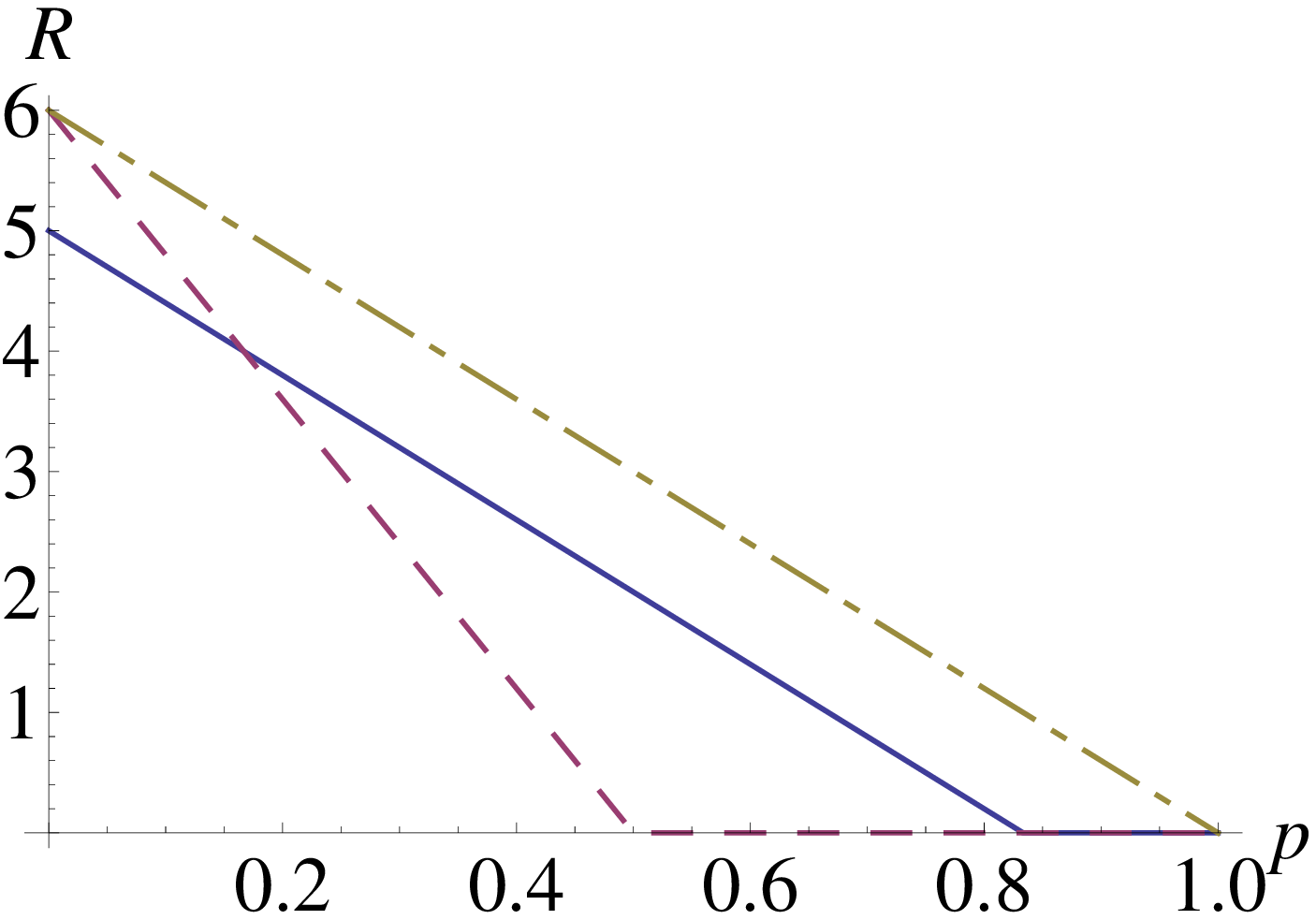}
\includegraphics[width=0.235\textwidth]{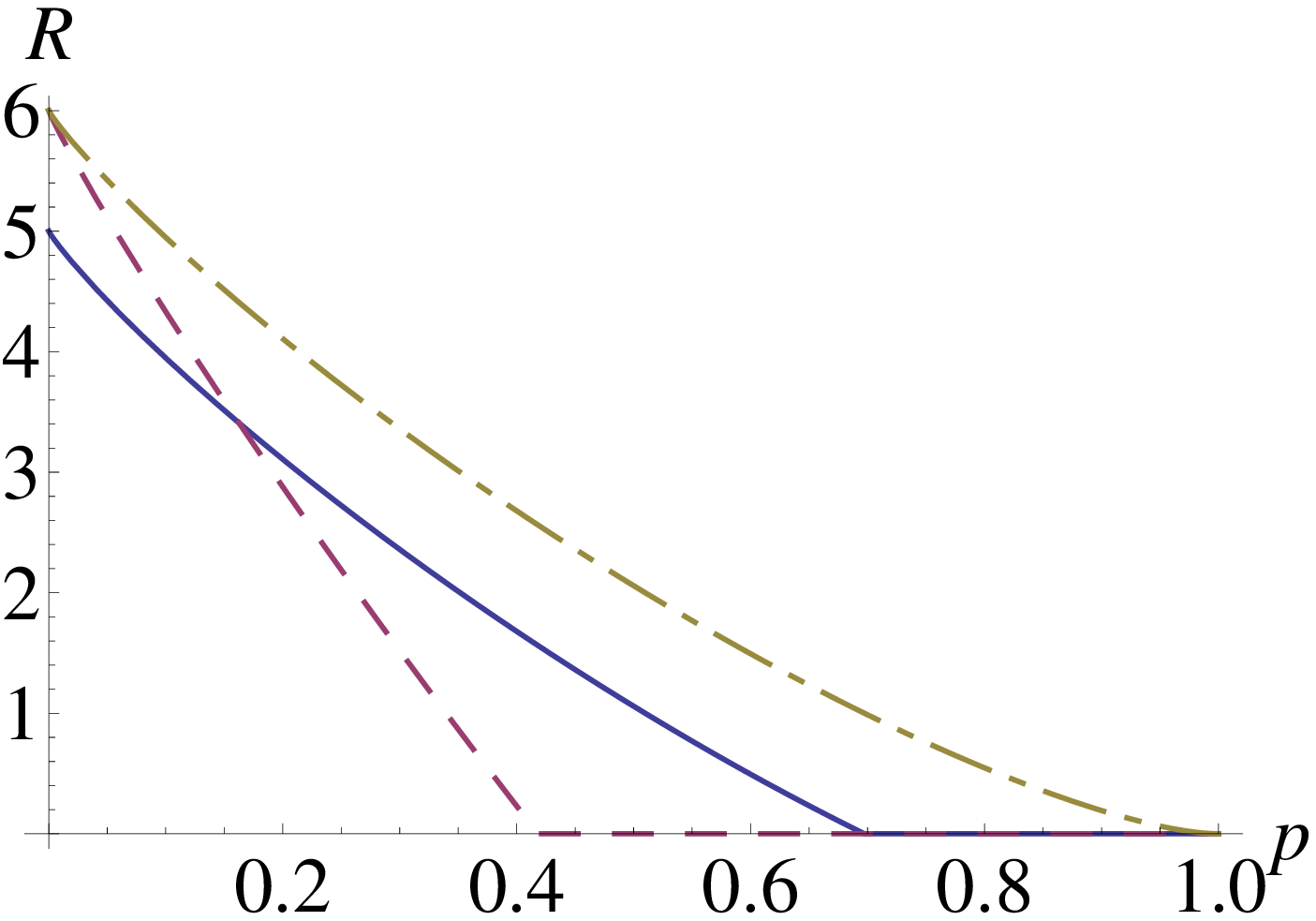}
\caption{
Comparison of several communication rates (in bits per channel use).
Left: asymptotic rates for the qudit erasure channel as a function
of the erasure probability $p$.
QDL secret key generation rate (solid line);
private capacity $P = (1-2p)\log{d}$ (dashed line);
classical capacity $C = (1-p)\log{d}$ (dash-dotted line).
Right: asymptotic rates for the qudit depolarizing channel as
function of the depolarizing probability $p$.
QDL secret key generation rate (solid line);
asymptotic secret key generation rate achieved by the protocol in~\cite{SS} 
(we notice incidentally that this rate achieves the Hashing bound) (dashed line);
classical capacity (dash-dotted line).
}
\label{fig:rates}
\end{figure}

Figure~\ref{fig:rates} shows a comparison of the secret key generation rates of our
protocol $R = C_\mathcal{N} - 1$ with the classical capacity and the private capacity (which equals
the secret key generation rate with the assistance of $1$-way public communication from
Alice to Bob) for the qudit erasure and depolarizing channels.

{\it Conclusions.--} According to the QDL effect, a large gap exists between two natural
security definitions, one related to the Holevo information and the other to the
accessible information (the difference between these two entropic quantities is known as quantum discord \cite{QD}). 
In this Letter we have shown that, if the latter criterion is assumed, one can generate
a secret key through a memoryless noisy channel at a rate as high as the classical capacity 
minus one bit, independently of the channel private capacity.
The price to pay for such a high rate of secret key generation is that the accessible
information criterion does not guarantee unconditional and composable security. 
Our protocol guarantees composable security under the assumption that Alice and Bob 
know that the coherence time of Eve's quantum memory is no larger than $\tau$.
Interestingly enough, the key generation rate is independent on the value of $\tau$, as long as 
Alice and Bob know this value (though large values of $\tau$ would make the protocol unpractical). 

One should also ensure that the QDL is robust under leakage
to Eve of a small fraction of the key or the message. 
Indeed, as a small key allows one to (un)lock a disproportionate amount of information, it could very
well happen that the leakage to Eve of a few bits may allow her to uncover a much larger portion
of the message. This problem has been recently addressed in~\cite{phase}, where it is shown that
there exist QDL protocols that can be made resilient to loss of a given amount of information 
by increasing the secret key consumption by a proportional amount. The conclusions of~\cite{phase}
may be straightforwardly generalized to the protocol discussed here, and hence 
applied to guarantee the robustness of our QDL protocol for noisy channels. 

The QDL states and unitaries in Eqs.~(\ref{FT}) and~(\ref{Uphase}) are particularly suitable for quantum optics
applications, where a qudit can be encoded by coherently splitting a single photon over $d$ modes
(e.g., path, temporal, linear momentum, orbital angular momentum) and then by applying
i.i.d.\ random phases to the different modes by modulating an array of phase shifters. 
For example, this kind of transformation can
be implemented by group velocity dispersion and our protocol can be realized by a simple modification
of standard $d$-dimensional quantum key distribution protocols, see e.g.,~\cite{Dirk}.
As discussed in~\cite{QEM} this requires passive linear optical transformations and photo detection. 
In the unary encoding of a single photon over $d$ modes, linear losses are modeled by a qudit erasure
channel, and the depolarizing channel model provides a standard benchmark for assessing the performance 
of quantum key distribution.
Different channel models reflect different collective attacks conducted by the eavesdropper. 
While the final key generation rate may depend on the channel model, the security of our
QDL protocol (which holds in the strong locking sense) does not depend on the details of the channel, and 
it also holds in the case of coherent attacks. 
Finally, let us remark that unlike previous QDL protocols the one presented here does not 
require $d$ to be arbitrarily large.
Instead, our protocol requires an increasing number of channel uses (as typical of i.i.d.~information theory)
while it is sufficient to assume $d \geq 3$.

{\it Acknowledgment.--} We are grateful to Fr\'ed\'eric Dupuis, 
Andreas Winter, and especially to Mark M. Wilde for helpful discussions and comments. 
This research was supported by the DARPA Quiness Program through 
U.S.~Army Research Office Grant No.~W31P4Q-12-1-0019.

%

%%%
\bigskip 

\appendix

\begin{center}
{\huge Supplemental Material}
\end{center}

\section{A quantum data locking protocol with separable codewords}\label{sec:noiseless}

Let us consider a $d$-dimensional Hilbert space endowed with an orthonormal basis 
$\{ |\omega\rangle \}_{\omega=1,\dots,d}$ and its Fourier-conjugate basis $\{ |m\rangle \}_{m=1,\dots,d}$, 
\begin{equation}\label{FT}
|m\rangle = \frac{1}{\sqrt{d}} \, \sum_{\omega=1}^d e^{i 2\pi m \omega/d} |\omega\rangle \, .
\end{equation}
Given a collection of $n$ qudit systems, we consider the product basis 
vectors $|\boldsymbol{m}\rangle = \otimes_{j=1}^n |m^j\rangle$. 
To encode the message $\boldsymbol{m} = (m^1,m^2,\dots,m^n)$, Alice prepares the state $|\boldsymbol{m}\rangle$.
Alice and Bob publicly agree on a subset of $K$ $n$-qudit unitaries
\begin{equation}
\mathcal{U}_k = \bigotimes_{j=1}^n U^j_k  \, ,
\end{equation}
for $k=1,2,\dots,K$, where the single-qudit unitary acting on the $j$-th system is of the form
\begin{equation}
U^j_k = \sum_{\omega^j=0}^{d-1} e^{i\theta^j_k(\omega^j)} |\omega^j\rangle \langle \omega^j| \, .
\end{equation}
According to the value of a pre-shared secret key, $k=1,2,\dots,K$, Alice and Bob privately agree on using
one of these unitaries for locking the codewords.
Alice applies the unitary transformation $\mathcal{U}_k$ on the $n$-qudit codeword, obtaining
\begin{eqnarray}
|\Psi_{\boldsymbol{m}k}\rangle & = & \mathcal{U}_k |\boldsymbol{m}\rangle = \bigotimes_{j=1}^n U^j_k |m^j\rangle \\
& = & \frac{1}{\sqrt{d^n}} \, \sum_{\boldsymbol{\omega}} e^{i\sum_{j=1}^n \left[ 2\pi m^j \omega^j/d + \theta^j_k(\omega^j) \right]} |\boldsymbol{\omega}\rangle \, ,
\label{ncw}
\end{eqnarray}
where 
$\boldsymbol{\omega} = (\omega^1,\omega^2,\dots,\omega^n)$ and
$|\boldsymbol{\omega}\rangle = \otimes_{j=1}^n |\omega^j\rangle$.
Notice that, for any given $k$, the vectors $\{ |\Psi_{\boldsymbol{m}}\rangle \}_{\boldsymbol{m}}$ define 
a new basis for the $n$-qudit system.

We consider the ``phase ensemble'' of qudit unitary transformations of the form:
\begin{equation}
U = \sum_{\omega=1}^d e^{i\theta(\omega)} |\omega\rangle \langle \omega| \, ,
\end{equation}
where the angles $\theta(\omega)$, for $\omega=1,\dots,d$, are $d$ i.i.d.\ random variables. 
We require that these variables are distributed in such a way
that $\mathbb{E}[e^{i\theta(\omega)}] = 0$.

Below we show that, if the unitaries $U^j_k$ are randomly chosen, identically and independently, from
the phase ensemble, then the data locking protocol will succeed with a probability arbitrary
close to $1$ if $n$ is large enough. In particular, this protocol requires a secret key of 
$\log{K} \simeq n$ bits, that is, the protocol consumes secret key at an asymptotic rate of $1$ bit per 
data-locked qudit.

%%%

\subsection{Some preliminary results}

To characterize our QDL protocol we will make use of two concentration inequalities.
The first one is the Maurer tail bound~\cite{AM-supp}:
\begin{theorem}\label{Maurer}
Let $\{ X_t \}_{t=1,\dots,T}$ be $T$ i.i.d.\ non-negative real-valued random variables, with $X_t \sim X$ and $\mathbb{E}[X],\mathbb{E}[X^2] < \infty$.
Then, for any $\tau > 0$ we have that
$$
Pr\left\{ \frac{1}{T}\sum_{t=1}^T X_t < \mathbb{E}[X] - \tau \right\} \leq \exp{\left( - \frac{T\tau^2}{2\mathbb{E}[X^2]} \right)} \, .
$$
\end{theorem}
($Pr\{ x \}$ denotes the probability that the proposition $x$ is true.)
The second one is the operator Chernoff bound~\cite{Chernoff-supp}:
\begin{theorem}\label{Chernoff}
Let $\{ X_t \}_{t=1,\dots,T}$ be $T$ i.i.d.\ random variables taking values in the algebra of hermitian operators in dimension $D$,
with $0 \leq X_t \leq \mathbb{I}$ and $\mathbb{E}[X_t] = \mu \mathbb{I}$ ($\mathbb{I}$ is the identity operator).
Then, for any $\tau > 0$ we have that
$$
Pr\left\{ \frac{1}{T}\sum_{t=1}^T X_t > (1+\tau)\mu\mathbb{I} \right\} \leq D \, \exp{\left( - \frac{T\tau^2\mu}{4\ln{2}} \right)} \, .
$$
\end{theorem}

%%%

For any given $d^n$-dimensional unit vector $|\Phi\rangle$, $\boldsymbol{m}$ and $k$, we define the quantity
\begin{equation}
q_{\boldsymbol{m}k}(\Phi) = |\langle \Phi | \Psi_{\boldsymbol{m}k} \rangle|^2 \, ,
\end{equation}
which is a function of the codeword $|\Psi_{\boldsymbol{m}k}\rangle$ defined by Eq.~(\ref{ncw}).
Notice that $q_{\boldsymbol{m}k}(\Phi)$ is a random variable for a random choice of the set of scrambling unitaries $\{\mathcal{U}_k\}_{k=1,\dots,K}$
To apply Theorems~\ref{Maurer} and~\ref{Chernoff}, we compute the first and second moments of $q_{\boldsymbol{m}k}(\Phi)$
with respect to the i.i.d.\ random unitaries sampled from the phase ensemble.
Putting $|\Phi\rangle = \sum_{\boldsymbol{\omega}} \Phi_{\boldsymbol{\omega}} |\boldsymbol{\omega}\rangle$, we have
\begin{widetext}
\begin{eqnarray}
\mathbb{E}[q_{\boldsymbol{m}k}(\Phi)] & = & \frac{1}{d^n} \sum_{\boldsymbol{\omega},\boldsymbol{\omega}'} \Phi^*_{\boldsymbol{\omega}} \Phi_{\boldsymbol{\omega}'} 
\, e^{i\sum_{j=1}^n2\pi m^j (\omega^j-{\omega'}^j)/d }
\, \mathbb{E}\left[ e^{i\sum_{j=1}^n \left[ \theta^j(\omega^j) - \theta^j({\omega'}^j)\right]} \right] \\
& = & \frac{1}{d^n} \sum_{\boldsymbol{\omega},\boldsymbol{\omega}'} \Phi^*_{\boldsymbol{\omega}} \Phi_{\boldsymbol{\omega}'} 
\, e^{i\sum_{j=1}^n2\pi m^j (\omega^j-{\omega'}^j)/d }
\, \prod_{j=1}^n \mathbb{E}\left[ e^{i \theta^j(\omega^j) - i \theta^j({\omega'}^j)} \right] \\
& = & \frac{1}{d^n} \sum_{\boldsymbol{\omega},\boldsymbol{\omega}'} \Phi^*_{\boldsymbol{\omega}} \Phi_{\boldsymbol{\omega}'} 
\, e^{i\sum_{j=1}^n2\pi m^j (\omega^j-{\omega'}^j)/d }
\prod_{j=1}^n \delta_{\omega^j {\omega'}^j} \\
& = & \frac{1}{d^n} \sum_{\boldsymbol{\omega},\boldsymbol{\omega}'} \Phi^*_{\boldsymbol{\omega}} \Phi_{\boldsymbol{\omega}'} 
\prod_{j=1}^n \delta_{\omega^j {\omega'}^j} = \frac{1}{d^n} \, , \label{1stm}
\end{eqnarray}
and
\begin{eqnarray}
\mathbb{E}[q_{\boldsymbol{m}k}(\Phi)^2] & = & \frac{1}{d^{2n}} \sum_{\boldsymbol{\omega},\boldsymbol{\omega}',\boldsymbol{\omega}'',\boldsymbol{\omega}'''} 
\Phi^*_{\boldsymbol{\omega}} \Phi_{\boldsymbol{\omega}'} \Phi^*_{\boldsymbol{\omega}''} \Phi_{\boldsymbol{\omega}'''} 
\, e^{i\sum_{j=1}^n2\pi m^j (\omega^j-{\omega'}^j+{\omega''}^j-{\omega'''}^j)/d } \nonumber \\ 
& \times & \mathbb{E}\left[ e^{i\sum_{j=1}^n \left[ \theta^j(\omega^j) - \theta^j({\omega'}^j) + \theta^j({\omega''}^j) - \theta^j({\omega'''}^j) \right]} \right] \\
& = & \frac{1}{d^{2n}} \sum_{\boldsymbol{\omega},\boldsymbol{\omega}',\boldsymbol{\omega}'',\boldsymbol{\omega}'''} 
\Phi^*_{\boldsymbol{\omega}} \Phi_{\boldsymbol{\omega}'} \Phi^*_{\boldsymbol{\omega}''} \Phi_{\boldsymbol{\omega}'''} 
\, e^{i\sum_{j=1}^n2\pi m^j (\omega^j-{\omega'}^j+{\omega''}^j-{\omega'''}^j)/d } \nonumber \\ 
& \times & \prod_{j=1}^n \mathbb{E}\left[ e^{i\theta^j(\omega^j) - i\theta^j({\omega'}^j) + i\theta^j({\omega''}^j) - i\theta^j({\omega'''}^j) } \right] \\
& = & \frac{1}{d^{2n}} \sum_{\boldsymbol{\omega},\boldsymbol{\omega}',\boldsymbol{\omega}'',\boldsymbol{\omega}'''} 
\Phi^*_{\boldsymbol{\omega}} \Phi_{\boldsymbol{\omega}'} \Phi^*_{\boldsymbol{\omega}''} \Phi_{\boldsymbol{\omega}'''} 
\, e^{i\sum_{j=1}^n2\pi m^j (\omega^j-{\omega'}^j+{\omega''}^j-{\omega'''}^j)/d } \nonumber \\ 
& \times & \prod_{j=1}^n \frac{ \delta_{\omega^j {\omega'}^j} \delta_{{\omega''}^j {\omega'''}^j} + \delta_{\omega^j {\omega'''}^j} \delta_{{\omega'}^j {\omega''}^j} }{ 1 + \delta_{\omega^j{\omega''}^j}} \\
& = & \frac{1}{d^{2n}} \sum_{\boldsymbol{\omega},\boldsymbol{\omega}',\boldsymbol{\omega}'',\boldsymbol{\omega}'''} 
\Phi^*_{\boldsymbol{\omega}} \Phi_{\boldsymbol{\omega}'} \Phi^*_{\boldsymbol{\omega}''} \Phi_{\boldsymbol{\omega}'''} 
\prod_{j=1}^n \frac{ \delta_{\omega^j {\omega'}^j} \delta_{{\omega''}^j {\omega'''}^j} + \delta_{\omega^j {\omega'''}^j} \delta_{{\omega'}^j {\omega''}^j} }{ 1 + \delta_{\omega^j{\omega''}^j}} \, .
\end{eqnarray}
\end{widetext}
One can show that (see Section~\ref{app:m2}):
\begin{equation}
\mathbb{E}[q_{\boldsymbol{m}k}(\Phi)^2] 
\leq \frac{2^n}{d^{2n}} \, .
\end{equation}

%%%

For any given $|\Phi\rangle$ and $\boldsymbol{m}$ we define the quantity $Q_{\boldsymbol{m}}(\Phi)$
by taking the average over $k$:
\begin{equation}
Q_{\boldsymbol{m}}(\Phi) = \frac{1}{K} \sum_{k=1}^K q_{\boldsymbol{m}k}(\Phi) \, .
\end{equation}
Notice that for $k \neq k'$, $|\Psi_{\boldsymbol{m}k}\rangle$ and $|\Psi_{\boldsymbol{m}k'}\rangle$
are statistically independent, and so are $q_{\boldsymbol{m}k}(\Phi)$ and $q_{\boldsymbol{m}k'}(\Phi)$.
We can hence apply Maurer's tail bound (Theorem~\ref{Maurer}). We obtain that for any given $|\Phi\rangle$ and $\boldsymbol{m}$:
\begin{equation}\label{small}
Pr\left\{ Q_{\boldsymbol{m}}(\Phi) < \frac{1-\epsilon}{d^n} \right\} \leq \exp{\left( - \frac{K \epsilon^2}{2^{n+1}} \right)} \, .
\end{equation}
We then apply the operator Chernoff bound (Theorem~\ref{Chernoff}) to the operators $|\Psi_{\boldsymbol{m}k}\rangle\langle\Psi_{\boldsymbol{m}k}|$.
Notice that $\mathbb{E}[q_{\boldsymbol{m}k}(\Phi)] = 1/d^{n}$ [Eq.~(\ref{1stm})] implies
\begin{equation}
\mathbb{E}[|\Psi_{\boldsymbol{m}k}\rangle\langle\Psi_{\boldsymbol{m}k}|] = \frac{\mathbb{I}}{d^n} \, .
\end{equation}
The operator Chernoff bound then yields that for any given $\boldsymbol{m}$
\begin{align}
& Pr\left\{ \frac{1}{K} \sum_{k=1}^K |\Psi_{\boldsymbol{m}k}\rangle\langle\Psi_{\boldsymbol{m}k}| > (1-\delta) \mathbb{I} \right\} \nonumber \\
& \leq d^n \exp{\left( -\frac{K (d^n(1-\delta)-1)^2}{d^n 4 \ln{2}} \right)} \\
& = d^n \exp{\left( -\frac{K d^n (1-\delta-1/d^n)^2}{4 \ln{2}} \right)} \, .
\end{align}
This result in turn implies that for any given $\boldsymbol{m}$
\begin{align}
& Pr\left\{ \max_{|\Phi\rangle} Q_{\boldsymbol{m}}(\Phi) > 1-\delta \right\} \nonumber \\
& \leq d^n \exp{\left( -\frac{K d^n (1-\delta-1/d^n)^2}{4 \ln{2}} \right)} \, . \label{large}
\end{align}

%%%

Finally, to optimize Eve's measurement on her share of the quantum state, we will make use of the notion of $\epsilon$-net.
An $\epsilon$-net is a finite set of unit vectors $\mathcal{N}_\epsilon = \{ |\Phi_i\rangle \}_i$
in a $D$-dimensional Hilbert space such that for any unit vector $|\Phi\rangle$ there exists 
$|\Phi_i\rangle \in \mathcal{N}_\epsilon$ such that
\begin{equation}\label{enet}
\| |\Phi\rangle\langle\Phi| - |\Phi_i\rangle\langle\Phi_i| \|_1 \leq \epsilon \, .
\end{equation}
As discussed in~\cite{CMP-supp} there exists an $\epsilon$-net such that $|\mathcal{N}_\epsilon| \leq (5/\epsilon)^{2D}$.

\subsection{Eve's accessible information}

In the strong locking scenario, we assume that Eve intercepts the whole train of qudit systems and measures them.
To evaluate the security of the QDL protocol according to the accessible information criterion, we show
that there exist choices of the scrambling unitaries $U^j_k$'s that guarantee Eve's accessible information to be 
arbitrarily small if $n$ is large enough.
To prove that we show that this property is true with a non-zero probability for a random choice of the
unitaries $U^j_k$'s.
The proof strategy is analogous to the one of~\cite{phase-supp} and is based on similar ideas already applied
to other QDL protocols~\cite{CMP-supp,Fawzi-supp}. 

Let Eve intercept and measure the train of $n$ qudits sent by Alice. A measurement is described by the POVM elements
$\{ \mu_i |\Phi_i\rangle\langle\Phi_i| \}_i$, where $\sum_i \mu_i = d^n$, $\mu_i >0$ and $|\Phi_i\rangle$ are unit
vectors (possibly entangled over the $n$ qudit systems).
Since Eve does not have access to the secret key, we have to compute the accessible information of the 
ensemble of states $\mathcal{E} = \{ p_{\boldsymbol{m}} , \frac{1}{K} \sum_{k=1}^K |\Psi_{\boldsymbol{m}k}\rangle \langle \Psi_{\boldsymbol{m}k} |\}$,
where $p_{\boldsymbol{m}}$ is the probability of the message $\boldsymbol{m}$.
For the sake of simplicity here we assume that all the messages have equal probability, that is, $p_{\boldsymbol{m}} = 1/d^n$
(the case of non-uniform distribution was considered in~\cite{Fawzi-supp,Dupuis-supp}).
A straightforward calculation then yields
\begin{equation}
I_{acc}\left( \mathcal{E} \right) = \log{d^n} - \min_{\{ \mu_i |\Phi_i\rangle\langle\Phi_i| \}} \sum_i \frac{\mu_i}{d^n} \, H[Q(\Phi_i)]  \, ,
\end{equation}
where $Q(\Phi)$ denotes the $d^n$-dimensional real vector with non-negative entries
\begin{equation}
Q_{\boldsymbol{m}}(\Phi) = \frac{1}{K} \sum_{k=1}^K |\langle \Phi | \Psi_{\boldsymbol{m}k} \rangle |^2 
= \frac{1}{K} \sum_{k=1}^K q_{\boldsymbol{m}k}(\Phi) \, ,
\end{equation}
and 
\begin{equation}
H[Q(\Phi)] = - \sum_{\boldsymbol{m}} Q_{\boldsymbol{m}}(\Phi) \log{Q_{\boldsymbol{m}}(\Phi)}
\end{equation}
is its Shannon entropy
(notice that $\sum_{\boldsymbol{m}} Q_{\boldsymbol{m}}(\Phi) = 1$).

Since $\sum_i \mu_i/d^n = 1$, the positive coefficients $\mu_i/d^n$ can be interpreted as probability
weights. We can then apply a standard convexity argument (the minimum is never larger than the average) 
to obtain an upper bound on Eve's accessible information:
\begin{equation}\label{Iaccnl}
I_{acc}\left( \mathcal{E} \right) \leq \log{d^n} - \min_{|\Phi\rangle} \, H[Q(\Phi)]  \, ,
\end{equation}
where the minimum is over all $n$-qudit unit vectors.
According to this expression, an upper bound on the accessible information follows
from a lower bound on the minimum Shannon entropy $\min_{|\Phi\rangle} \, H[Q(\Phi)]$.

%%%

In order to prove that $I_{acc}\left( \mathcal{E} \right) \leq \epsilon\log{d^n}$, 
we need to show that $\min_{|\Phi\rangle} \, H[Q(\Phi)] \geq \left( 1 - \epsilon \right) \log{d^n}$.
To do that, for any $\epsilon > 0$ and $d^n$ and $K$ large enough we bound the probability that 
\begin{equation}
- Q_{\boldsymbol{m}}(\Phi) \log{Q_{\boldsymbol{m}}(\Phi)} < \eta\left( \frac{1-\epsilon}{d^n} \right) \, ,
\end{equation}
where $\eta(x) := -x \log{x}$.
This is obtained by bounding the probability that 
either $Q_{\boldsymbol{m}}(\Phi) < \lambda_-$
or $Q_{\boldsymbol{m}}(\Phi) > \lambda_+$, 
where 
$\lambda_- = (1-\epsilon)/d^n$
and 
$\lambda_+ = 1 - \eta\left( \frac{1-\epsilon}{d^n} \right) + O\left(\eta\left( \frac{1-\epsilon}{d^n} \right)\right)$.
Notice that for $d^n$ sufficiently large and/or $\epsilon$ sufficiently small we have
$\lambda_+ \geq 1 - 2\eta\left( \frac{1-\epsilon}{d^n} \right)$.

%%%

From Eq.~(\ref{large}) and applying the union bound we obtain 
\begin{widetext}
\begin{eqnarray}
Pr\left\{ \max_{|\Phi\rangle,\boldsymbol{m}} Q_{\boldsymbol{m}}(\Phi) > \lambda_+ \right\} 
& \leq & Pr\left\{ \max_{|\Phi\rangle,\boldsymbol{m}} Q_{\boldsymbol{m}}(\Phi) > 1 - 2\eta\left( \frac{1-\epsilon}{d^n} \right) \right\} \\
& \leq & d^n Pr\left\{ \max_{|\Phi\rangle} Q_{\boldsymbol{m}}(\Phi) > 1 - 2\eta\left( \frac{1-\epsilon}{d^n} \right) \right\} \\
& \leq & d^{2n} \exp{\left( -\frac{K d^n (1-2\eta\left( \frac{1-\epsilon}{d^n} \right)-1/d^n)^2}{4 \ln{2}} \right)} \\
& \leq & \exp{\left( \ln{d^{2n}} -\frac{K d^n (1-2\eta\left( \frac{1-\epsilon}{d^n} \right)-1/d^n)^2}{4 \ln{2}} \right)} \\
& \leq & \exp{\left( \ln{d^{2n}} -\frac{K d^n (1-4\eta\left( \frac{1-\epsilon}{d^n} \right)-2/d^n)}{4 \ln{2}} \right)} \\
& \leq & \exp{\left( \ln{d^{2n}} -\frac{K d^n (1-6\eta\left( \frac{1-\epsilon}{d^n}\right))}{4 \ln{2}} \right)} =: p_+ \, , \label{Plarge}
\end{eqnarray}
where we have also used the fact that $\frac{1}{d^n} < \eta\left( \frac{1-\epsilon}{d^n}\right)$ for $n$ large enough.
This probability vanishes exponentially with $d^{n}$ provided that 
$K > \frac{\ln{d^{2n}}}{d^n} \frac{4 \ln{2}}{1-6\eta[(1-\epsilon)/d^n]}$.

%%%

Then, for any given $|\Phi\rangle$ we use Eq.~(\ref{small}) and apply again the union bound to obtain
\begin{eqnarray}
Pr\left\{ \exists \boldsymbol{m}_1,\dots,\boldsymbol{m}_\ell \, \, | \, \, \forall i \, \, Q_{\boldsymbol{m}_i}(\Phi) < \lambda_- \right\}
& = & Pr\left\{ \exists \boldsymbol{m}_1,\dots,\boldsymbol{m}_\ell \, \, | \, \, \forall i \, \, Q_{\boldsymbol{m}_i}(\Phi) < \frac{1-\epsilon}{d^n} \right\} \\
& \leq & { d^n \choose \ell} \left( Pr\left\{ Q_{\boldsymbol{m}}(\Phi) < \frac{1-\epsilon}{d^n} \right\} \right)^\ell \\
& \leq & { d^n \choose \ell} \exp{\left( - \frac{\ell K \epsilon^2}{2^{n+1}}\right)} \\
& \leq & (d^n)^\ell \,  \exp{\left( - \frac{\ell K \epsilon^2}{2^{n+1}}\right)} \\
& = & \exp{\left( \ell \ln{d^n} - \frac{\ell K \epsilon^2}{2^{n+1}}\right)} \, .
\end{eqnarray}
\end{widetext}
Putting $\ell = \epsilon d^n$ we have
\begin{align}
& Pr\left\{ \exists \boldsymbol{m}_1,\dots,\boldsymbol{m}_\ell \, \, | \, \, \forall i \, \, Q_{\boldsymbol{m}_i}(\Phi) < \lambda_- \right\} \nonumber \\
& \leq \exp{\left[ - d^n \left( \frac{K \epsilon^3}{2^{n+1}} - \epsilon \ln{d^n} \right) \right]} =: p_- \, . \label{Psmall}
\end{align}
Notice that this probability is also exponentially small in $d^n$, provided that
$K > 2^{n+1} \epsilon^{-2} \ln{d^n}$.

Inequality~(\ref{Psmall}) implies that with probability greater than $1-p_-$
there are at least $d^n-\ell = (1-\epsilon) d^n$ values of $\boldsymbol{m}$ such that
$Q_{\boldsymbol{m}}(\Phi) > \lambda_-$. 
Also, according to Eq.~(\ref{Plarge}), with probability at least equal to $1-p_+$
all the $Q_{\boldsymbol{m}}(\Phi)$'s are larger than $\lambda_+$.
Putting these results together we have that
\begin{align}
H[Q(\Phi)] & > - (1-\epsilon) d^n \left( \frac{1-\epsilon}{d^n} \log{\frac{1-\epsilon}{d^n}} \right) \\
& = - (1-\epsilon)^2 \log{\frac{1-\epsilon}{d^n}} \\
& > (1-2\epsilon) \log{d^n} - (1-2\epsilon) \log{(1-\epsilon)} \\
& > (1-2\epsilon) \log{d^n} 
\end{align}
with a probability at least equal to $1 - p_- - p_+$.
For $d^n$ large enough this probability is larger than $1 - 2 p_-$.

%%%

The last step is to introduce an $\epsilon$-net $\mathcal{N}_\epsilon = \{ |\Phi_i\rangle \}_i$.
Let us recall that the $\epsilon$-net can be chosen to contain less than $(5/\epsilon)^{2d^n}$ elements.
We can hence apply the union bound to obtain:
\begin{align}
& Pr\left\{ \min_{|\Phi_i\rangle \in \mathcal{N}_\epsilon} H[Q(\Phi_i)] < (1-2\epsilon) \log{d^n}\right\} \nonumber \\ 
& \leq (5/\epsilon)^{2d^n} \, 2 p_- \\ 
& = 2 (5/\epsilon)^{2d^n} \exp{\left[ - d^n \left( \frac{K \epsilon^3}{2^{n+1}} - \epsilon \ln{d^n} \right) \right]} \\
& = 2 \exp{\left[ - d^n \left( \frac{K \epsilon^3}{2^{n+1}} - \epsilon \ln{d^n} - 2\log{\frac{5}{\epsilon}} \right) \right]} \, .
\end{align}
Finally, we have to replace the minimum over vectors in the $\epsilon$-net with a minimum over all unit vectors. 
An application of the Fannes inequality~\cite{FA-supp} yields (see also~\cite{CMP-supp})
\begin{equation}
\left| \min_{|\Phi\rangle} H[Q(\Phi)] - \min_{|\Phi_i\rangle \in \mathcal{N}_\epsilon} H[Q(\Phi_i)] \right| \leq \epsilon \log{d^n} + \eta(\epsilon) \, .
\end{equation}
This result implies
\begin{align}
& Pr\left\{ \min_{|\Phi\rangle} H[Q(\Phi)] < (1-3\epsilon) \log{d^n} - \eta(\epsilon) \right\} \nonumber \\
& \leq 2 \exp{\left[ - d^n \left( \frac{K \epsilon^3}{2^{n+1}} - \epsilon \ln{d^n} - 2\log{\frac{5}{\epsilon}} \right) \right]} \, ,
\end{align}
that is,
\begin{align}
& Pr\left\{ \max_{|\Phi\rangle} I_{acc} > 3\epsilon \log{d^n} + \eta(\epsilon) \right\} \nonumber \\
& \leq 2 \exp{\left[ - d^n \left( \frac{K \epsilon^3}{2^{n+1}} - \epsilon \ln{d^n} - 2\log{\frac{5}{\epsilon}} \right) \right]} \,.
\end{align}
Such a probability is bounded away from one (and goes to zero exponentially in $d^n$) provided 
\begin{equation}
K > 2^{n+1} \,  \left( \frac{1}{\epsilon^2} \ln{d^n} + \frac{2}{\epsilon^3}\log{\frac{5}{\epsilon}} \right) \, .
\end{equation}

In conclusion, we have proven that there exist QDL codes allowing Alice and Bob to lock
data through $n$ uses of a noiseless memoryless qudit channel in such a way that Eve's accessible information
is $I_\mathrm{acc}(\mathcal{E}) = O\left( \epsilon \log{d^n} \right)$.
These codes are defined by codewords that are separable among different channel uses.
The rate of locked communication is of $\log{d}$ bits per channel use and require the pre-shared
secret key to be consumed at an asymptotic rate of $\lim_{n \to \infty} \frac{1}{n}\log{K} = 1$ bit per channel use.
Notice that we can put $\epsilon = 2^{-n^c}$, with any positive $c<1$ and still lock data with a
secret key consumption rate of $1$ bit independently of $d$.

\section{Second moment of $q_{\boldsymbol{m}k}(\Phi)$}\label{app:m2}

Let us put
\begin{equation}
g_{\boldsymbol{\omega}\boldsymbol{\omega}'\boldsymbol{\omega}''\boldsymbol{\omega}'''} 
= \prod_{j=1}^n \frac{ \delta_{\omega^j {\omega'}^j} \delta_{{\omega''}^j {\omega'''}^j} + \delta_{\omega^j {\omega'''}^j} \delta_{{\omega'}^j {\omega''}^j} }{ 1 + \delta_{\omega^j{\omega''}^j}} \, .
\end{equation}
Notice $g_{\boldsymbol{\omega}\boldsymbol{\omega}'\boldsymbol{\omega}''\boldsymbol{\omega}'''}$ takes values in $\{ 0, 1\}$
and that the number of times it is equal to $1$ is $(2d^2-d)^n$.
Then we have (summation over repeated indexes is assumed) 
\begin{eqnarray}
f(\Phi) = d^{2n} \, \mathbb{E}[q_{mk}(\Phi)^2] = 
g_{\boldsymbol{\omega}\boldsymbol{\omega}'\boldsymbol{\omega}''\boldsymbol{\omega}'''} 
\Phi^*_{\boldsymbol{\omega}} \Phi_{\boldsymbol{\omega}'} \Phi^*_{\boldsymbol{\omega}''} \Phi_{\boldsymbol{\omega}'''} \, .
\end{eqnarray}

Let us define the $d^{2n} \times d^{2n}$ matrix $G$ with entries:
\begin{equation}
G^{(n)}_{\boldsymbol{\omega}\boldsymbol{\omega}'',\boldsymbol{\omega}'\boldsymbol{\omega}'''} := g_{\boldsymbol{\omega}\boldsymbol{\omega}'\boldsymbol{\omega}''\boldsymbol{\omega}'''} \, ,
\end{equation}
where $\boldsymbol{\omega}\boldsymbol{\omega}''$ and $\boldsymbol{\omega}'\boldsymbol{\omega}'''$
are respectively row and column indexes. Then we have
\begin{equation}
f(\Phi) \leq \| G^{(n)} \|_\infty \, ,
\end{equation}
where $\| G^{(n)} \|_\infty$ denotes the maximum eigenvalue of the matrix $G^{(n)}$.
We then notice that $G^{(n)} = G^{\otimes n}$, where $G$ is the $d^2 \times d^2$ matrix with entries
(no summation over repeated indexes)
\begin{align}
& G_{\omega\omega'',\omega'\omega'''} = \frac{ \delta_{\omega {\omega'}} \delta_{{\omega''} {\omega'''}} + \delta_{\omega{\omega'''}} \delta_{{\omega'}{\omega''}} }{ 1 + \delta_{\omega{\omega''}}} \nonumber \\ 
& = \delta_{\omega {\omega'}} \delta_{{\omega''} {\omega'''}} + \delta_{\omega{\omega'''}} \delta_{{\omega'}{\omega''}} - \delta_{\omega\omega'}\delta_{\omega'\omega''}\delta_{\omega''\omega'''} \, .
\end{align}
We have
\begin{equation}
G = I + S - P \leq I + S \, ,
\end{equation}
where $I$ is the $d^2 \times d^2$ identity matrix, $S$ is the swap matrix, and $P$ is the positive semidefinite
matrix with entries $P_{\omega\omega'',\omega'\omega'''} = \delta_{\omega\omega'}\delta_{\omega'\omega''}\delta_{\omega''\omega'''}$.

Since $I$ and $S$ are unitary (and hermitian) their eigenvalues are not greater than $1$,
which implies $\| G \|_\infty \leq 2$ and $\| G^{(n)} \|_\infty \leq 2^n$.
In conclusion we obtain $f(\Phi) \leq 2^n$ and $\mathbb{E}[q_{mk}(\Phi)^2] \leq 2^n/d^{2n}$.


\begin{thebibliography}{99}
\bibitem{NOTA02}
Vice versa, the Pinsker inequality (see e.g.,~\cite{Pisker} and references therein) implies that 
if $I_\mathrm{acc}\left( \mathcal{E} \right) \leq \epsilon$, then the norm in Eq.~(\ref{vecnorm}) is smaller than
$\sqrt{(2 \ln{2})\epsilon}$ \, .
\bibitem{NOTA-Holevo}
The Holevo information does not suffer from this dependence on external variables, such as the timing
of the measurement. This is the reason why the latter is the preferred and widely
accepted security criterion for quantum cryptography.
\bibitem{NOTA-nopa}
We remark that this setting corresponds to the one detailed in Section 4.1 of Ref.~\cite{errcorr}, 
where Eve is unable to eavesdrop on any information from the quantum channel (due to the QDL
effect in our case) but transmission errors may occur in the communication from Alice to Bob
(due to the noise introduced by the channel).
\bibitem{NOTA01}
For instance, the angles $\theta(\omega)$ can be uniformly distributed in $[0,2\pi[$,
or assume the binary values $\theta(\omega) \in \{ 0 , \pi \}$ with equal probabilities.
\bibitem{NOTA-proof}
The proof strategy is analogous to the one of~\cite{phase} and is based on similar ideas already applied
to other QDL protocols~\cite{CMP,Fawzi}.
\bibitem{NOTA-key}
It could, in principle, be possible to improve this bound.
\end{thebibliography}
\end{document}